\documentclass[11pt, oneside]{article}   	
\usepackage{geometry}                		
\usepackage{graphicx}				
\usepackage{amssymb}
\usepackage{url}

\title{\MakeLowercase{partitura}: A Python Package for Handling \\ Symbolic Musical Data\footnote{This preprint is a slightly updated version of the extended abstract presented at the Late Breaking/Demo Session of the 20th International Society for Music Information Retrieval Conference (ISMIR 2019), Delft, The Netherlands.}}
\author{Maarten Grachten$^1$\hspace{1cm} Carlos Cancino-Chac\'{o}n$^2$ \hspace{1cm} Thassilo Gadermaier$^2$ \\ \\
$^1$ Independent Researcher, Barcelona, Spain\\
$^2$  Institute of Computational Perception, Johannes Kepler University Linz, Austria\\
{\tt\small maarten.grachten@gmail, carlos\_eduardo.cancino\_chacon@jku.at, thassilo.gadermaier@jku.at}
}
\date{}							

\begin{document}
\maketitle

\begin{abstract}
This demo paper introduces \emph{partitura}, a Python package for handling symbolic musical information.
The principal aim of this  package is to handle richly structured musical information as conveyed by modern staff music notation. 
It provides a much wider range of possibilities to deal with music than the more reductive (but very common) piano roll-oriented approach inspired by the MIDI standard.
The package is an open source project and is available on GitHub.\footnote{\url{https://www.github.com/CPJKU/partitura}}
\end{abstract}
\section{Introduction}
In this work we present \emph{partitura}, a Python package for handling the symbolic musical information that is conveyed by modern staff notation.
The package was born out of a need to process richly structured musical information in a less reductive way than the piano roll representation that is very common in MIR, in which a score is represented as a list of timed pitch events.
Although there are certainly valid use cases for piano roll representations of music, we believe that some musical tasks can be more effectively addressed based on a richer data representation.
Computational modeling of musical expression is one such task.

Musical scores contain a variety of musically relevant information that is typically not present in a piano roll representation, including but not limited to pitch spelling, metrical structure, phrasing, voicing, articulation, tempo, dynamics, and musical form.
A challenge when dealing with this information is that it requires more complex data structures than the matrix structure typically used to represent piano rolls.
The \emph{partitura} package uses the notion of a \emph{timeline} to express the temporal scope of the elements in a score, such as notes, rests, slurs, measures, time and key signatures, and performance directions.
Elements may contain references to each other.
For example, a slur contains references to the starting and ending note of the slur.
This approach is further illustrated below.

The package supports exporting and importing musical scores to and from files in \emph{MusicXML} and \emph{MIDI} format.
Although the MIDI format in itself does not retain much of the musical information that \emph{partitura} intends to capture, the package includes proven algorithms for pitch spelling, voice estimation, and key estimation (see below), to reconstruct some of that information.

In relation to the well-known \emph{music21}\footnote{\url{https://web.mit.edu/music21/}} Python package it should be noted that the aims of \emph{partitura} are more modest.
Whereas \emph{music21} provides a toolkit for computer-aided musicology---including functionality like visualization and searching corpora---\emph{partitura} aims to facilitate processing musical information in Python.
It roughly follows \emph{MusicXML} in terms of musical entities, but as opposed to \emph{MusicXML}, where time is largely implicit, \emph{partitura} takes a strongly time-oriented approach.
This approach allows for extracting local musical contexts in full detail, but makes it equally straightforward to extract subsets of information from the score as a whole.

The rest of this document is structured as follows:
Section \ref{sec:partitura} discusses how musical scores are defined in \emph{partitura}.
Section \ref{sec:music_analysis_tools} presents the music analysis tools included in the package.
Finally, Section \ref{sec:conclusions} concludes this paper.

\section{Partitura}\label{sec:partitura}

In \emph{partitura} a score is defined at the highest level by one or more \textit{Part} objects, possibly grouped by \emph{PartGroup} objects. Parts are typically associated with instruments, and each part may have one or more staves.
Each Part contains a \textit{TimeLine} object that encapsulates a sequence of \textit{TimePoint} objects, each denoting a temporal position in the score (in an attribute $t$).
A musical element such as a \emph{Note} is added to the TimeLine by registering it with the TimePoints corresponding to its start and end positions.
A particularly important element is the \emph{Divisions} element, because it specifies the relation between the time interval \texttt{tp2.t} - \texttt{tp1.t} between two timepoints \texttt{tp1} and \texttt{tp1}, and the duration of a quarter note.

\begin{figure}[t]
\centering
\includegraphics[width=\linewidth]{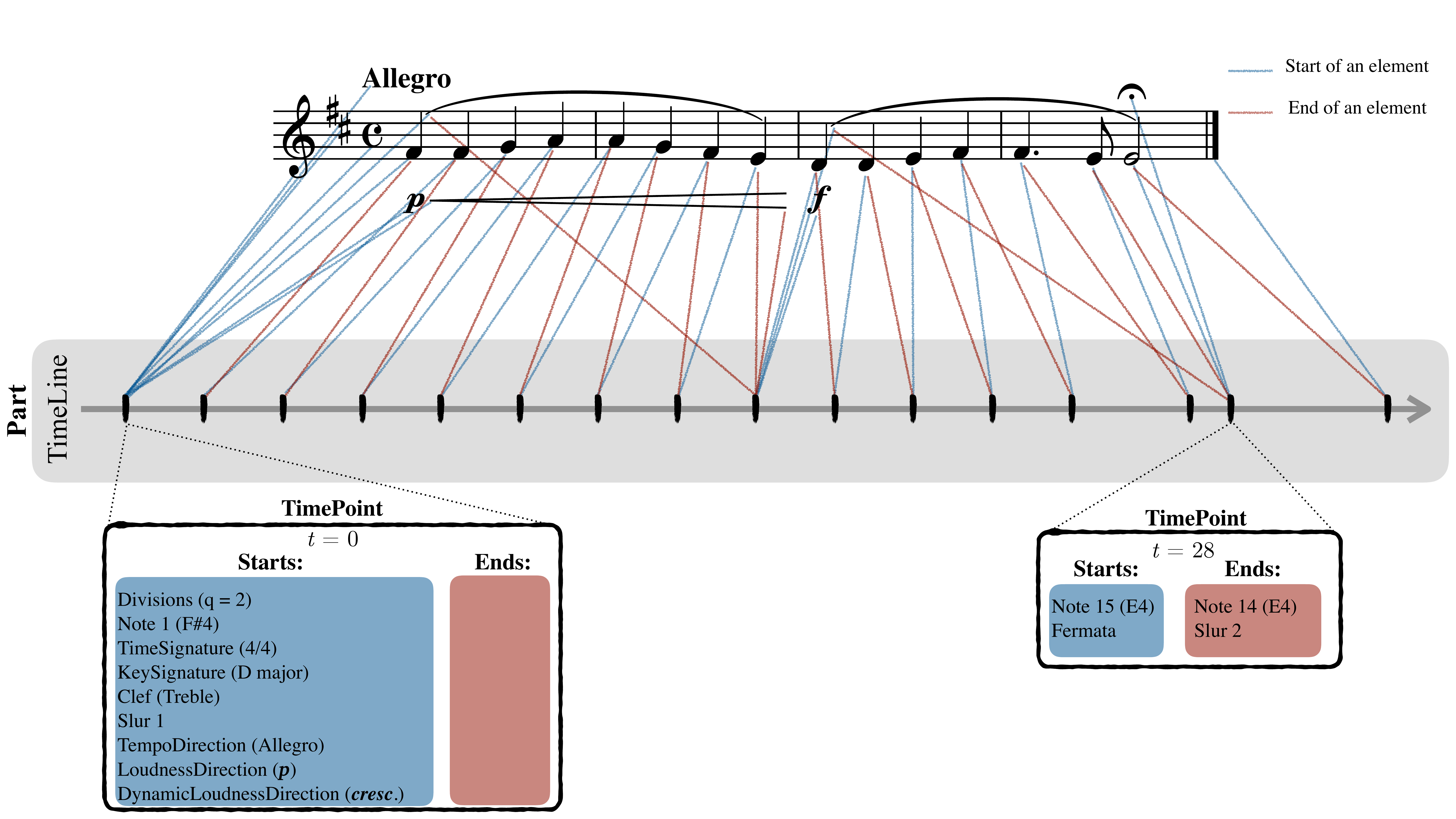}
\caption{Schematic representation of a Part object: A part contains a TimeLine object, which holds TimePoints (i.e., pegs that fix score elements in time). The blue lines represent the starting times of the objects in the score and the red lines represent the end times. }
\label{fig:timeline_example}
\end{figure}

Figure \ref{fig:timeline_example} shows a schematic representation of a Part object and its components.

\section{Music Analysis Tools}\label{sec:music_analysis_tools}

As mentioned above, \textit{partitura} includes some tools for music analysis which are intended to fill in missing information with plausible values, for instance when loading a score from a MIDI file.
For estimating the key signature of a piece, we use the Krumhansl--Shepard key identification algorithm~\cite{krumhansl1990}.
We include an implementation of the \emph{ps13s1} algorithm~\cite{Meredith:ps13} for estimating pitch spelling.
For estimating voice information, we use \emph{VoSA}~\cite{Chew2004}, a contig mapping approach for voice separation in polyphonic music.
To our knowledge, this is the first publicly available Python implementation of ps13s1 and VoSA.

\section{Conclusions and Future Work}\label{sec:conclusions}

The package is available on \emph{GitHub}, with documentation available at \emph{readthedocs.org}\footnote{\url{https://partitura.readthedocs.io/en/latest/index.html}}.
Future work will include support of the MEI format\footnote{\url{https://music-encoding.org}} and the match format which is used to encode performance-to-score alignments.

\section*{Acknowledgments} 
This research has received funding from the European Research Council (ERC) under the European Union's Horizon 2020 research and innovation programme under grant agreement No. 670035 (project ``Con Espressione") and the Austrian Science Fund (FWF) under grant P 29840-G26 (project ``Computer-assisted Analysis of Herbert von Karajan's Musical Conducting Style'').

\bibliographystyle{apalike}
\bibliography{partitura}

\end{document}